\documentstyle[12pt,psfig]{article}
\begin{document}
\begin{titlepage}
\begin{flushright}
{\setlength{\baselineskip}{0.18in}
{\normalsize
METU-PHYS-HEP98/02\\
}}
\end{flushright}
\vskip 2cm
\begin{center}
{\Large\bf
Implications of $\Delta \rho$ and CHARM II data for $Z^{\prime}$ 
Physics }

\vskip 1cm
{\large
D. A. Demir\footnote{e-mail:ddemir@heraklit.physics.metu.edu.tr\\ Present Address:
ICTP, Trieste, Italy}

\vskip 0.5cm
{\setlength{\baselineskip}{0.18in}
{\normalsize\it Middle East Technical University, Department of Physics,
06531, Ankara, Turkey\\} }}
\end{center}
\vskip .5cm
\begin{abstract}
We discuss the constraints on the $Z^{\prime}$ model parameters coming 
from $Z$- pole and low-energy $\nu_{\mu}-e$ scattering data in the frame 
work of GUT- motivated models. We find that when the coupling constant of 
the extra $U(1)$ is small (large) the parameter space is mainly 
determined by the $Z$-pole (the low-energy $\nu_{\mu}-e$ scattering) data. 
\end{abstract}
\end{titlepage}
\newpage
\section{Introduction}
Deviations of the precisely measured electroweak observables from the 
predictions of the SM \cite{pdg-lang} enable one to search for the 
existence of new physics. Although there are various possibilities for 
extending the SM, one of the simplest and well-motivated extension is 
the addition of an extra $U(1)$ to its $SU(3)_{C}\times SU(2)_{L}\times 
U(1)_{Y}$ gauge structure \cite{rizzo}. The Z-pole and neutral current 
data can be used to search for and set limits on the existence of the 
$Z^{\prime}$ models \cite{lang-luo-mann}. The Z-pole observables are 
sensitive only to the $Z-Z^{\prime}$ mixing angle due to the 
modifications in the vector and axial couplings of the fermions to the 
Z-boson. On the other hand, off-Z-pole $e^{+}e^{-}$ data, and low energy 
$\nu_{\mu}-e$, $\nu$-hadron and parity-violating $e$-hadron 
scattering experiments are sensitive to both $Z-Z^{\prime}$ mixing and 
$Z'$ mass \cite{lang-luo-mann}. Therefore, accompanying the Z-pole data,  
the low-energy observables are particularly useful in establishing the 
constraints on the $Z'$ physics. For example, atomic parity violation 
experiments (particularly for cesium) has already been analyzed in this 
direction \cite{marciano-rosner,mahant}.
     
In this work we investigate the implications of $\Delta \rho$ and CHARM 
II data \cite{charm} on the $Z^{\prime}$ models by comparing them with 
the SM predictions \cite{novikov,ewwg,taliev}. In the analysis, we 
illustrate two GUT-motivated models and reach conclusions of general 
applicability.

In Sec. 2 we give the relevant formulae for the effective vector and 
axial couplings in the frame work of the $Z^{\prime}$ models. Moreover, 
we list SM expressions and derscribe the GUT-motivated $Z^{\prime}$  
models that will be the subject matter of the analysis.

In Sec. 3 we give a detailed numerical analysis of the appropriate 
parameter space by taking into account both points implied by the CHARM 
II data \cite{charm}.

In Sec. 4 we conclude the work and remark on the interplay between 
low-energy and $Z$- pole determinations of the parameter space.

\section{$\nu_{\mu}-e$ Scattering in $Z^{\prime}$ Models}
In addition to the usual SM gauge group, we consider an additional 
Abelian group $U(1)_{Y'}$ with coupling constant $g_{Y'}$, under 
which the left-handed lepton doublets and right-handed charged leptons 
have charges $Q_L$ and $Q_E$, respectively. The neutral vector 
bosons $Z$ of $SU(2)_{L}\times U(1)_{Y}$ and $Z^{\prime}$ of  
$U(1)_{Y'}$ mix with each other. The mass eigenstate vector bosons 
$Z_{1}$ and $Z_{2}$ can be obtained after diagonalizing the $Z-Z'$ 
mass-squared matrix:
\begin{eqnarray}
Z_1&=&\cos\theta Z+\sin\theta Z^{\prime}\\
Z_2&=&-\sin\theta Z + \cos\theta Z^{\prime}
\end{eqnarray}
which define the $Z-Z^{\prime}$ mixing angle $\theta$. The $Z^{\prime}$ 
effects can show up in various physical quantities whose comparison with 
the SM predictions allow us to constrain $Z^{\prime}$ parameters in a 
model independent way. In general, $Z'$ effects can be tracked by their 
apprearence in four distinct physical quantities. Firstly, due to mixing, 
$Z_1$ is lighter than the canonical $Z$, so that the $\rho$ parameter 
predicted by $Z_1$ mass is larger than the one predicted by $Z$ mass. 
Secondly, coupling of $Z_1$ to fermions differs from those of $Z$ due to 
the mixing in (1). Thirdly, the $Z^{\prime}$ exchange will modify the 
neutral current amplitudes as an explicit function of the $Z'$ mass. 
Finally, $\sin^{2}\theta_{W}$ differs from the SM prediction because of 
the $Z^{\prime}$ contribution to the $\rho$ parameter. In what follows we 
shall discuss all of these effects in the framework of the $\mu_{\mu}-e$ 
scattering.

The low-energy four-fermion effective lagrangian relevant to the 
$\nu_{\mu}-e$ scattering can be written as 
\begin{eqnarray}
-L^{v_{\mu}-e}=\frac{G_{F}}{\sqrt{2}}\bar{\nu}_{\mu}\gamma^{\alpha}
(1-\gamma^{5})\nu_{\mu}\,
\bar{e}\gamma_{\alpha}(g_{V}^{\nu_{\mu}e}-g_{A}^{\nu_{\mu}e}\gamma^{5})e
\end{eqnarray} 
which defines the effective vector coupling $g_{V}^{\nu_{\mu}e}$ and axial 
coupling $g_{A}^{\nu_{\mu}e}$. $\nu_{\mu}-e$ scattering is a pure neutral 
current process mediated by both $Z_1$ and $Z_2$. Thus 
$g_{V}^{\nu_{\mu}e}$ and $g_{A}^{\nu_{\mu}e}$ carry the $Z^{\prime}$ 
effects through the contribution of the intermediate vector bosons, and 
the modification of the fermion-vector boson couplings due to the mixing.
In the $Z^{\prime}$ model mentioned-above these couplings are given by
\begin{eqnarray}
g_{V}^{\nu_{\mu}e}&=&\rho_{1}\{\tilde{g}_{V}^{\nu_{\mu}e}\cos^{2}\theta+ 
\lambda(1+r+2\tilde{g}_{V}^{\nu_{\mu}e})\sin 2\theta + 
8\lambda^{2}(1+r)\sin^{2}\theta\}\nonumber\\ &+& 
\rho_{2}\{\tilde{g}_{V}^{\nu_{\mu}e}\sin^{2}\theta-
\lambda(1+r+2\tilde{g}_{V}^{\nu_{\mu}e})\sin 2\theta + 
8\lambda^{2}(1+r)\cos^{2}\theta\}\\ 
g_{A}^{\nu_{\mu}e}&=&\rho_{1}\{\tilde{g}_{A}^{\nu_{\mu}e}\cos^{2}\theta+
\lambda(1-r+2\tilde{g}_{A}^{\nu_{\mu}e})\sin 2\theta +
8\lambda^{2}(1-r)\sin^{2}\theta\}\nonumber\\ &+&
\rho_{2}\{\tilde{g}_{A}^{\nu_{\mu}e}\sin^{2}\theta-
\lambda(1-r+2\tilde{g}_{A}^{\nu_{\mu}e})\sin 2\theta +
8\lambda^{2}(1-r)\cos^{2}\theta\}
\end{eqnarray}
Here $\rho_1$ and $\rho_2$ are the $\rho$ parameters of $Z_1$ and $Z_2$:
\begin{eqnarray}
\rho_1&=&\frac{1+\eta^{2} \tan^{2}\theta}{1+\tan^{2}\theta}\\
\rho_2&=&\frac{1+\eta^{2} \tan^{2}\theta}{\eta^{2}(1+\tan^{2}\theta)}
\end{eqnarray}
where $\eta = M_{Z_{2}}/M_{Z_{1}}$. Next $\lambda = 
(g_{Y'}Q_{L})/(2 G)$, $r=Q_{E}/Q_{L}$, and 
$G=\sqrt{g_{2}^{2}+g_{Y}^{2}}$. Finally, the tilded vector and axial 
couplings are given by
\begin{eqnarray}
\tilde{g}_{V}^{\nu_{\mu}e}&=&(g_{V}^{\nu_{\mu}e})_{SM}-2(\frac{ 
s^{2}_{W} c^{2}_{W}}{c^{2}_{W}-s^{2}_{W}})_{SM}\Delta \rho\\
\tilde{g}_{A}^{\nu_{\mu}e}&=&(g_{A}^{\nu_{\mu}e})_{SM}
\end{eqnarray}
where the subscript SM refers to the SM value of the associated quantity. 
Here $\Delta \rho$ is the deviation of the $\rho_1$ from its $SM$ value:
\begin{eqnarray}
\Delta \rho = \rho_1 -1 = (\eta^2 -1)\sin^{2}\theta
\end{eqnarray}
It is the $Z^{\prime}$ contribution to $s_{W}^{2}$ that causes a finite 
difference between $\tilde{g}_{V}^{\nu_{\mu}e}$ and 
$(g_{V}^{\nu_{\mu}e})_{SM}$. As we see $Z^{\prime}$ contribution to 
$s_{W}^{2}$ is proportional to $\Delta \rho$. 

We identify $g_{V}^{\nu_{\mu}e}$ in (4) and  $g_{A}^{\nu_{\mu}e}$ in (5) 
with the experimental result, and require the $Z^{\prime}$ contributions at 
their right hand sides to close the gap between experimental result and 
the SM determination. We use the SM results evaluated for $M_{H}=M_{Z}$ so 
that there is no contribution to a particular observable from the Higgs 
loop. This is a convenient approach as we have a much more complicated 
Higgs sector in $Z^{\prime}$ models (See, for example, \cite{us} and 
references therein) with increased number of scalars whose contributions 
to a particular observable differ from the one in SM. Moreover, the 
tree-level $Z^{\prime}$ contributions in (4) and (5) do already yield the 
relevant parameters of a $Z^{\prime}$ model to  be fixed at the leading 
order. Namely, we assume loop corrections to (4) and (5) coming from the 
$Z'$ sector negligably small.

Although we treat $Z^{\prime}$ effects at the tree- level, we take all 
the loop contributions into account when analyzing the SM contribution. 
One of the basic SM parameters entering (4) and (5) is the weak mixing 
angle $s_{W}^{2}$ whose meaning is to be fixed against the ambiguities 
coming from the renormalization scheme and scale. As was discussed in 
\cite{pdg-lang} in detail, the value of  $s_{W}^{2}$ extracted from 
$M_{Z}$ in $\bar{M}\bar{S}$ scheme at the scale $M_{Z}$ is less 
sensitive to $m_t$ and most types of the new physics compared to the 
others. Following, \cite{pdg-lang}, we adopt this definition of 
$s_{W}^{2}$ to be denoted by $\hat{s}^{2}_{Z}$ from now on. Consistent 
with this choice for $s_{W}^{2}$, the SM expressions for vector and axial 
couplings in (8) and (9) are given by 
\begin{eqnarray}
(g_{V}^{\nu_{\mu}e})_{SM} &=&\rho_{\nu e}(-1/2+2\hat{\kappa}_{\nu 
e}\hat{s}^{2}_{Z})\\
(g_{A}^{\nu_{\mu}e})_{SM} &=&\rho_{\nu e}(-1/2)
\end{eqnarray}
whose numerical values can be found in \cite{pdg-lang}.

In the analysis below we shall restrict $\eta$ and $\theta$ by requiring 
(4) and (5) be satisfied within the experimental error bounds. In 
addition to this, we take into account the constraints coming from $Z$ - 
pole data  by requiring $\Delta \rho $ (10) be in the band induced by 
the present world average for $\rho_1$ \cite{pdg-lang}: 
\begin{eqnarray}
0.999\leq \rho_1 \leq 1.0006
\end{eqnarray}
which puts an additional restriction on the allowed $\eta - \theta$ space. 
$\Delta \rho$ calculated from (13) includes $m_t$ effects already; however, 
Higgs effects are discarded by taking $M_H=M_Z$ as mentioned before. 
Although $\Delta \rho=\rho_1-1$ takes both negative and positive values, 
$Z'$ models require it be positive as $M_{Z_{1}}$ is less than $M_Z$ due 
to mixing.

The lepton charges $Q_L$ and $Q_E$ and gauge coupling $g_{Y'}$ are 
model dependent. In literature there are various $Z^{\prime}$ models 
\cite{mahant} with definite predictions for charges and gauge coupling. 
In this work we shall illustrate the constraints implied by CHARM II 
results by considering two typical GUT- motivated $Z^{\prime}$ models. 
Obviously, when choosing the models one should discard those that predict 
a right- handed neutrino as the form of the effective lagrangian (3) 
suggests. 

The first model is $I$ model coming from $E_{6}$ \cite{rizzo,dark}, and 
we call it Model ${\cal{B}}$. In this model $Q_{L}=-2 Q_{E}=1/\sqrt{10}$ 
and $g_{Y'}\simeq 0.8 g_{2}$. This is a typical $E_{6}$ -inspired model 
in which $U(1)_{Y'}$ charges are linear combinations of the $U(1)$ 
groups $E_{6}/SO(10)$ and $SO(10)/SU(5)$. Depending on the details of 
the symmetry breaking, $g_{Y'}$ differs from $0.8 g_{2}$ by a factor 
around unity.
 
The second model we analyze is the one following from the breaking of 
the flipped $SU(5)\times U(1)$ when the Higgs fields reside in the  
$(27+\bar{27})$ dimensional representation of $E(6)$ \cite{rizzo,dark}. In 
this model $Q_{L}=Q_{E}=1/2$ and $g_{Y'}$ is as in the first model. We 
call this model as Model ${\cal{A}}$.

In the notation of equations (4) and (5), we have $r_{{\cal{A}}}=1$, 
$r_{{\cal{B}}}=-1/2$, and $\lambda_{{\cal{B}}}\approx 0.63\, 
\lambda_{{\cal{A}}}$. Since $r_{{\cal{A}}}=1$, $g_{A}^{\nu_{\mu}e}$ 
feels $Z^{\prime}$ parameters only through $\lambda \sin 2\theta$ type 
terms (see (5)) which is a weaker dependence than that of 
$g_{V}^{\nu_{\mu}e}$. Thus, in Model ${\cal{A}}$ parameter space is 
expected to be constrained mainly by $\Delta \rho$ and 
$g_{V}^{\nu_{\mu}e}$. 
 
In Model ${\cal{B}}$, as $r_{{\cal{B}}}=-1/2$, dependence of 
$g_{A}^{\nu_{\mu}e}$ on $Z^{\prime}$ parameters is amplified compared to 
$g_{V}^{\nu_{\mu}e}$. Thus, one expects $\Delta \rho$ and
$g_{A}^{\nu_{\mu}e}$ mainly determine the appropriate parameter space . 

Finally, one can also predict the relative magnitudes of the parameter 
spaces in two models. Since $\lambda_{{\cal{B}}}\approx 0.63\,
\lambda_{{\cal{A}}}$, it is clear that in Model ${\cal{B}}$ one needs a 
wider parameter space than in Model ${\cal{A}}$ to close the gap between 
experiment and the SM. In the next section we shall quantify these 
qualitative arguments.

\section{CHARM II Results and $Z^{\prime}$ Physics}
CHARM II Collaboration \cite{charm} has measured $s_{W}^{2}$, 
$g_{V}^{\nu_{\mu}e}$ and $g_{A}^{\nu_{\mu}e}$ from the neutrino-electron 
scattering events using $\nu$ and $\bar{\nu}$ beams with $E_{\nu}\simeq 
25.7 \, GeV$. On $g_{V}-g_{A}$ plane there are two candidate points  
coming from electron-neutrino scattering data. However LEP results for
$A_{FB}(e^{+}e^{-}\rightarrow e^{+}e^{-})$ \cite{lep} prefer one of them  
at which $g_{V}^{\nu_{\mu}e}= -0.035\pm 0.017$ and $g_{A}^{\nu_{\mu}e}= 
-0.503\pm 0.017$ \cite{pdg-lang}. This solution, which we call point 
(I), is close to the SM prediction so that it can be taken as a 
confirmation of the standard electroweak theory. The other point which
is discarded by $e^{+}e^{-}$ data has approximately $g_{V}^{\nu_{\mu}e} 
\leftrightarrow g_{A}^{\nu_{\mu}e}$. This solution, which we call point (II),
is clearly far from being predictable by the SM. As noted in  
\cite{pdg-lang}, point (I) is choosen by assuming that the $Z$-pole data 
is dominated by a single $Z$. In multi $Z$ models, like the $Z^{\prime}$ 
models under concern, it would be convenient to discuss the implications 
of both points for $Z^{\prime}$ physics on equal footing. Below we first 
analyze the allowed parameter space on $\eta - \theta $ plane for the 
point (I), next we turn to the discussion of point (II). 
\subsection{Allowed $\eta - \theta$ values for point (I)}
Let us start discussing the implications of equations (4), (5), and (10) 
with a rough analysis. Equations (4) and (5) express experimental 
values of $g_{V}^{\nu_{\mu}e}$ and $g_{A}^{\nu_{\mu}e}$ in terms of the 
parameters of the extended model at hand. Equation (10), on the  other 
hand, represents the restriction on the modification of  
$\rho$ parameter due to $Z^{\prime}$ effects. As given in (13) 
$\Delta \rho$ (10) is restricted in a rather narrow error band.
As is seen from equations (4)-(9), $g_{V}^{\nu_{\mu}e}$ and 
$g_{A}^{\nu_{\mu}e}$ explicitly depend on $\Delta \rho$, unlike the atomic 
weak charge for which $\Delta \rho$ dependence almost cancel \cite{mahant}.
   
For a rough analysis of the parameter space at the point (I), one can 
neglect $\Delta \rho$ all together. Then $\tilde{g}_{V}^{\nu_{\mu}e}$ (8) 
reduces to the SM expression. Moreover, $\rho_1\rightarrow 1$, and this 
requires both $\eta^{2} \tan^{2}\theta$ and $\tan^{2}\theta$ be small, which 
would be satisfied only by a small enough $\theta$. That 
$g_{V}^{\nu_{\mu}e}$ and $g_{A}^{\nu_{\mu}e}$ are close to their SM 
values also confirms the need for a small $\theta$, as this reduces the 
$\rho_1$ dependent terms in (4) and (5) essentially to their SM expressions.
In this limit, $\rho_2\sim 1/\eta$, and for $\rho_2 \cos^{2}\theta$ type 
terms in (4) and (5) be suppressed one needs a large $\eta$, or 
equivalently, heavy enough $Z_2$. After making these rough observations 
we now turn to an exact numerical analysis of the allowed $\eta-\theta$ 
region for point (I), which will be seen to imply constraints beyond these 
expectations as one considers the nonvanishing values of the $\rho$ 
parameter.
 
In the numerical analysis below we first express, via (10), $\eta$ in 
terms of $\theta$ for a given value of $\Delta \rho$. Then we insert this 
solution to the expressions for $g_{V}^{\nu_{\mu}e}$ and 
$g_{A}^{\nu_{\mu}e}$ in (4) and (5). After expressing (4) and (5) in 
terms of $\theta$ and $\Delta \rho$ in this way we let $\theta$ vary from 
zero to higher values, meanwhile, $\Delta \rho$ wanders in its 
phenomenologically allowed range of values determined by (13). Under the 
variation of these parameters we pick up those points for which (4) and 
(5) do remain in their allowed range of values. 

We first analyze Model ${\cal{A}}$ mentioned in the last section. In this 
model $g_{A}^{\nu_{\mu}e}$ has a weaker dependence on the $Z^{\prime}$ 
parameters compared to $g_{V}^{\nu_{\mu}e}$ since $r=1$. Fig. 1 depicts the 
allowed parameter space for which (4), (5) and (10) satisfy the existing 
phenomenological bounds. As we see from Fig. 1 for small $\theta$, $\eta$ 
wanders in a rather wide range of values. For example, for 
$\theta=10^{-4}\pi $, $\eta$ varies from 8 to 80 which implies $0.7\, TeV 
\stackrel{<}{\sim} M_{Z_{2}} \stackrel{<}{\sim} 7\, TeV$. As $\theta$ 
increases, not only the upper bound but also the lower bound on $\eta$ 
decreases. For example, for $\theta \approx 0.0041$, $\eta$ ranges from 
1.3 to 6, that is,  $120\, GeV \stackrel{<}{\sim} M_{Z_{2}} 
\stackrel{<}{\sim} 550\, GeV$ therefore $Z_{2}$ is considerably light in 
this case. Moreover, as $\theta$ increases further, the 
allowed range of $\eta$ gets thinner and thinner and $M_{Z_{2}}$ settles 
approximately to  $120 \, GeV$. This goes on until $\theta=0.0286$ at 
which $g_{V}^{\nu_{\mu}e}$ hits its lower bound, and there are no 
($\theta, \eta$) pairs satisfying (4), (5) and (10) beyond this 
point. In sum, one concludes that for Model ${\cal{A}}$, 
$\theta_{max}=0.0286$ and  $M_{Z_{2}}^{min}\approx 120\, GeV$. For 
negative $\theta$ values, graph is approximately mirror symmetric of 
Fig.1 with respect to $\eta$ axis. $\theta$ takes its minimum value of 
$\theta_{min}=-0.236$ at which $g_{V}^{\nu_{\mu}e}$ reaches its upper bound.
Close to $\theta_{min}$, $\eta$ drops to $\sim 1.1$, yielding a light 
$Z_{2}$.
  
Now we analyze Model ${\cal{B}}$ to indentify the appropriate 
region it implies in the ($\theta, \eta$) plane. As we recall from the 
previous section, unlike Model ${\cal{A}}$, in this model both 
$g_{A}^{\nu_{\mu}e}$ is more sensitive to the $Z^{\prime}$ parameters 
than $g_{V}^{\nu_{\mu}e}$. Fig.2 depicts the appropriate ($\theta, 
\eta$) pairs for which (4), (5) and (10) reside in their allowed band of 
values. Here we show only $\theta \geq 0.2$ part of the entire 
$\theta$ range to magnify the region of maximum $\theta$ value. $\theta 
\leq 0.2$ part behaves similarly to Fig.1. As we see from Fig. 2, the 
allowed range of $\theta$ is rather wide compared to  Model ${\cal{A}}$.
This is what was expected by the discussions at the end of the last 
section. To be quantitave let us discuss some special points on Fig. 2. 
Here $\theta_{max}= 0.3896$ at which $g_{A}^{\nu_{\mu}e}$ hits its upper 
bound so that there is no acceptable ($\theta, \eta$) values beyond this 
point. Consistent with the comperatively large values of $\theta$,  
$\eta$ takes its smallest allowed value, that is, it takes the value of 
unity for every  $\theta$ allowed, including the $\theta \leq 0.2$ 
part not shown on Fig. 2. For negative $\theta$ values the allowed region 
is approximately symmetric to Fig. 2 with respect to $\eta$ axis and 
extends up to -1.533 at which $g_{A}^{\nu_{\mu}e}$ reaches its upper 
bound. Differently than the positive $\theta$ range, here $\eta$ does not 
take the value of unity altough it remains close to unity, lowering as 
low as 1.005, for $\theta$ values near $\theta_{min}=-1.533$. 

Athough  $\eta=1$ was never realized for Model ${\cal{A}}$, it is the 
case here. This shows that for Model ${\cal{B}}$ the envelope of the 
allowed ($\theta, \eta$) values is strictly determined by $\Delta \rho$ 
in (10), as $\eta=1$ is consistent with $\Delta \rho$ constraint. For 
Model  ${\cal{A}}$, however, $\Delta \rho$ has not the full control over 
$\eta$ values for a given $\theta$, other parameters, 
$g_{V}^{\nu_{\mu}e}$ and $g_{A}^{\nu_{\mu}e}$, constrain the total  
$\eta$ range.  
  
For both models ${\cal{A}}$ and ${\cal{B}}$ we have a wider range of  
allowed negative $\theta$ values which stems from the fact that negative 
$\theta$ changes the sign of the cross terms in (4) and (5), and thus, 
lowers the value of the right hand side compared to the positive 
$\theta$ case. This then requires a wider range of $\theta$ values. 
This widening of the allowed negative $\theta$ interval is also seen in 
the APV determination of \cite{mahant}. However, one notes that, in 
general, the allowed ranges of $\theta$ is much wider than those of the 
APV determination. Nevertheless, a direct comparison is not possible, 
because in \cite{mahant}, experimental data are taken with an error of 
two standard deviations, and top mass is assigned a much lower value of 
$\sim 90\, GeV$ than the present analysis.  
  
We conclude this subsection by noting that the allowed ranges of the 
mixing angle and $Z_{2}$ mass depend on the value of the normalized 
$U(1)_{Y'}$ coupling $\lambda$. If it is small, mainly the $Z$-pole 
observable $\Delta \rho$ takes control of the $Z^{\prime}$ parameter 
space, and the low-energy observables  $g_{V}^{\nu_{\mu}e}$ and 
$g_{A}^{\nu_{\mu}e}$ do not hit their phenomenological bounds until the 
mixing angle takes relatively large values. On the other hand, when 
$\lambda$ is large, the low-energy observables become more decisive on 
the $Z^{\prime}$ parameter space. Therefore, one rhoughly concludes that 
when the coupling constant of the extra $U(1)$ is large (small) compared 
to the weak coupling low-energy observables  ($Z$- pole observables) 
determine the allowed parameter space.
\subsection{Allowed $\eta - \theta$ values for point (II)}
In Sec. 3.1 we have discussed the implications of CHARM II candidate, 
point (I), for the $Z^{\prime}$ parameters. The other candidate, point 
(II), discarded by the $e^{+}e^{-}$ data, deserves also a discussion 
because it may be of interest in $Z^{\prime}$ models in which off- $Z$- 
pole data may be explained by the contribution  of the $Z^{\prime}$
propagator effects. Before an attempt to confirm the $e^{+}e^{-}$ data, 
one should first prove the existence of points on ($\theta, \eta$) plane 
solving (4), (5) and (10) simultaneously. The transition from the 
point (I) to point (II) involves the exchange $g_{V}^{\nu_{\mu}e} 
\leftrightarrow g_{A}^{\nu_{\mu}e}$. Thus, it is convenient to work with 
$\Sigma^{\nu_{\mu}e}=g_{V}^{\nu_{\mu}e} + g_{A}^{\nu_{\mu}e}$ and 
$\Delta^{\nu_{\mu}e}=g_{V}^{\nu_{\mu}e} - g_{A}^{\nu_{\mu}e}$ which are 
symmmetric and antisymmetric under $g_{V}^{\nu_{\mu}e}\leftrightarrow 
g_{A}^{\nu_{\mu}e}$, respectively. After interchanging the left hand 
sides of (4) and (5), $\Sigma^{\nu_{\mu}e}$ and $\Delta^{\nu_{\mu}e}$ 
can be expressed in terms of the corresponding SM ones and the other 
parameters. However, an attempt to find a consistent solution of 
$\Sigma^{\nu_{\mu}e}$,  $\Delta^{\nu_{\mu}e}$ and $\Delta \rho$ for a 
given model parameters runs into difficulty immediately. This can be 
seen as follows. $\Sigma^{\nu_{\mu}e}$ is symmetric under 
$g_{V}^{\nu_{\mu}e} \leftrightarrow g_{A}^{\nu_{\mu}e}$ so it points to 
the ($\theta, \eta$) regions we discussed in Sec. 3.1. However, 
$\Delta^{\nu_{\mu}e}$ is antisymmetric under  $g_{V}^{\nu_{\mu}e} 
\leftrightarrow g_{A}^{\nu_{\mu}e}$  and requiers a completely different 
($\theta, \eta$) set for being satisfied. This is due to the fact that 
$\Delta^{\nu_{\mu}e}$ and the corresponding SM quantity $\Delta^{SM}$ 
are not close to each other as $\Sigma^{\nu_{\mu}e}$ and $\Sigma^{SM}$ 
do. In fact, a numerical analysis of not only the models we 
have illustrated in Sec. 3.1 but also the ones listed in \cite{mahant} 
confirms this observation. It is interesting that the present 
analysis also prefers the CHARM II result for point (I).
\section{Conclusions}       
In this work we have analyzed the low energy $\nu_{\mu}-e$ scattering 
and $Z$ -pole data against a general Abelian extension of the SM. In 
our analysis we have used the results of CHARM II Collaboration 
\cite{charm} which reported two candidate points for the effective 
vector and axial couplings. We have considered both points in our 
analysis and found that the CHARM II candidate discarded by the 
$e^{+}e^{-}$ data does not lead  a consistent solution for the GUT 
-motivated models considered. The other point, which is in the close 
vicinity of the SM predictions for  effective vector and 
axial couplings, leads solution spaces much wider than the APV 
determination of \cite{mahant}. The  analysis concludes that when the 
gauge coupling associated to the extra $U(1)$ is small (large) compared 
to the weak coupling parameter space is determined mainly by $\rho$ 
-parameter constraint (low- energy neutral current data).

\newpage
\begin{figure}   
\vspace{5.0cm}
\end{figure}
\begin{figure}
\vspace{12.0cm} 
    \includegraphics{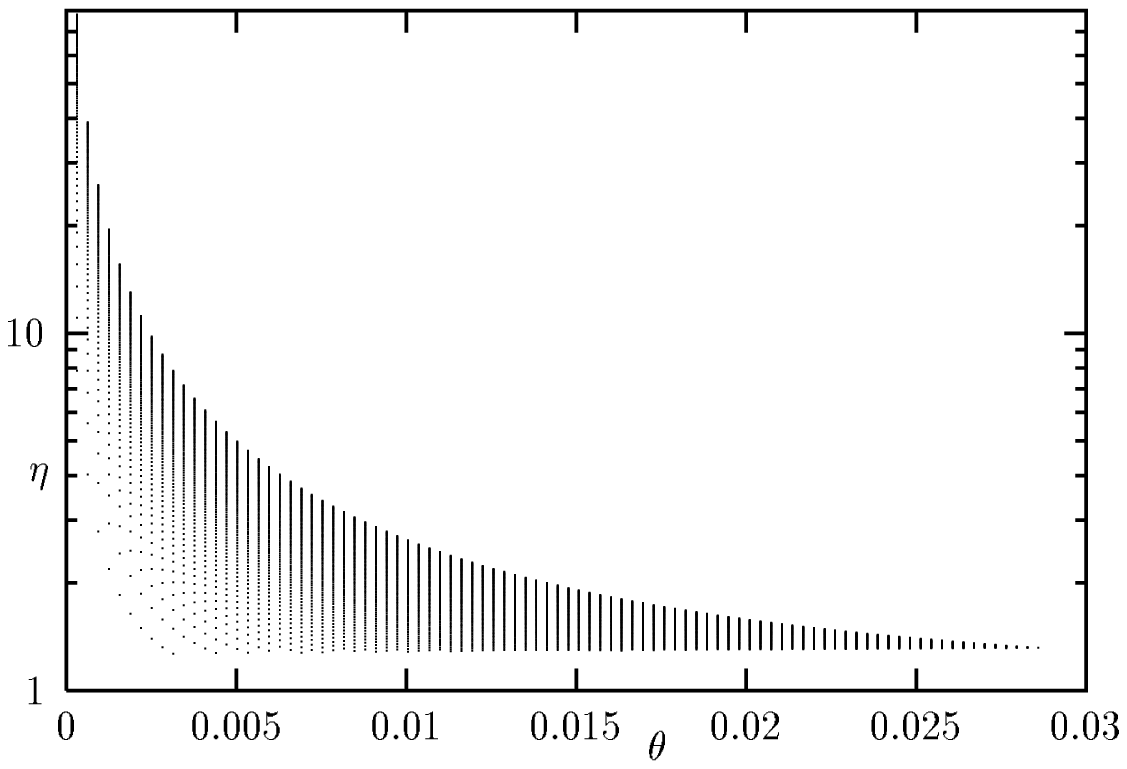}
    \vspace{-11.0cm}
\vspace{0.0cm}
\mbox{ \hspace{0.2cm} \large{\bf Figure 1: The ($\eta , \theta$) values 
satisfying (4), (5) and}}
\mbox{ \hspace{0.2cm} \large{\bf (10) for Model ${\cal{A}}$.}}
\end{figure}
\newpage
\begin{figure}
\vspace{12.0cm}
\end{figure}
\begin{figure}
\vspace{12.0cm}
    \includegraphics{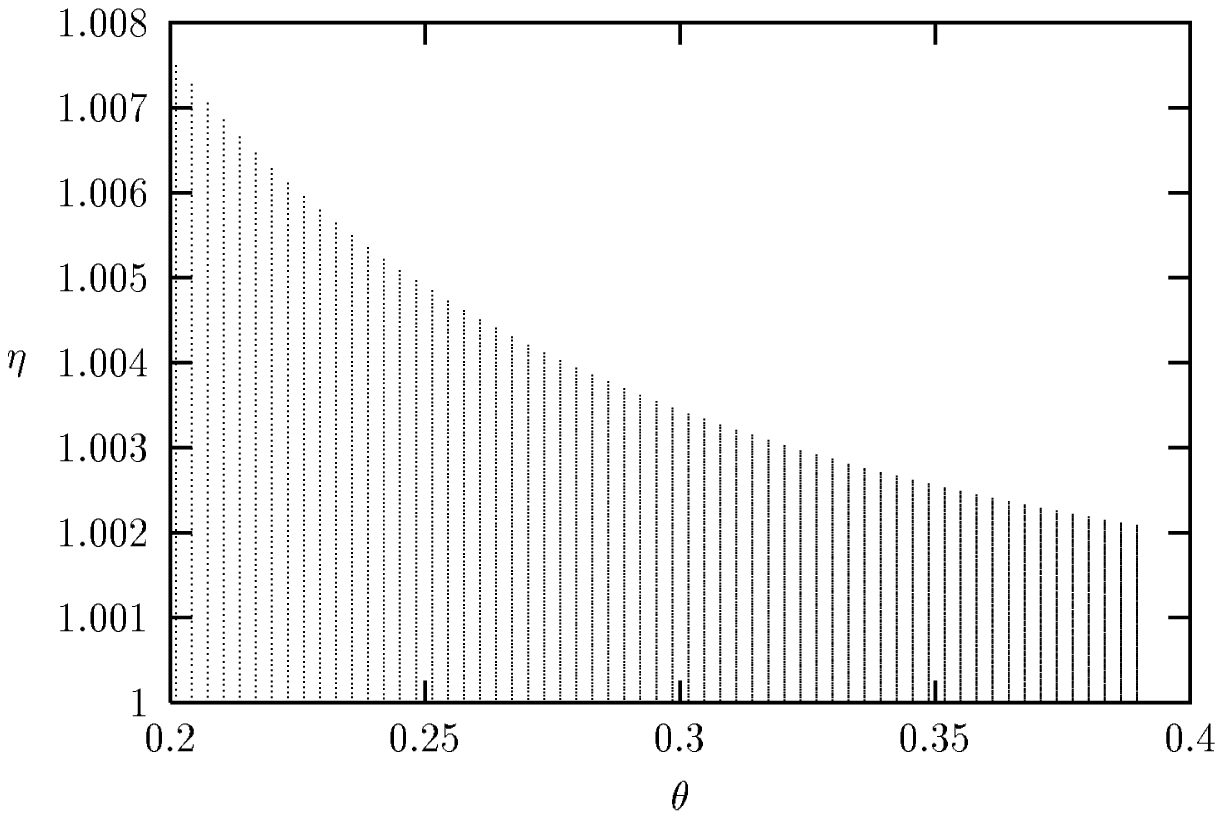}
    \vspace{-11.0cm}
\vspace{0.0cm}
\mbox{ \hspace{0.2cm} \large{\bf Figure 2: The ($\eta , \theta$) values
satisfying (4), (5) and}}
\mbox{ \hspace{0.2cm} \large{\bf (10) for Model ${\cal{B}}$.}}
\end{figure}
\end{document}